# Toward Sustainable Networking: Storage Area Networks with Network Coding

Ulric J. Ferner, Muriel Médard, and Emina Soljanin

*Abstract*—This manuscript provides a model to characterize the energy savings of network coded storage (NCS) in storage area networks (SANs). We consider blocking probability of drives as our measure of performance. A mapping technique to analyze SANs as independent $M/G/K/K$ queues is presented, and blocking probabilities for uncoded storage schemes and NCS are derived and compared. We show that coding operates differently than the amalgamation of file chunks and energy savings are shown to scale well with striping number. We illustrate that for enterprise-level SANs energy savings of 20–50% can be realized.

## I. Introduction

CURRENT projections indicate that the worldwide data center (DC) industry will require a quadrupling of capacity by the year 2020 [1], primarily through increased demand for high-definition video streaming. In addition to requiring significant financial investments, worldwide DC capacities are approaching a scale in which their energy consumption and carbon footprint is significant [2]. approaching the scale of the worldwide airline industry [1], [3], [4].

In this paper, we consider the energy efficiency of individual storage area networks that make up DCs. Storage area networks (SANs) are designed so that large numbers of content consumers can be serviced concurrently, while the average quality of the user experience is maintained. Specifically, the probability that a piece of content is unavailable to any user, and there being an interruption during consumption, is kept small. To achieve these small blocking probabilities, individual content is replicated on multiple drives [5]. This replication increases the chance that, if a server or drive with access to target content is unavailable, then another copy of the same file on a different drive can be read instead. Modern content replication strategies are designed to help SANs service millions of video requests in parallel [6]. Examples of services that use such scalable systems include YouTube [7] and Hulu [8].

The goal of this paper is to characterize the financial and energy savings of NCS in a single SAN. The key contributions include:

- We provide a $M/G/K/K$ model for SANs;

This material is based upon work supported by Alcatel-Lucent under award #4800484399 and the Jonathan Whitney MIT fellowship. U.J. Ferner and M. Médard are with the Research Laboratory for Electronics (RLE), Massachusetts Institute of Technology, Room 36-512, 77 Massachusetts Avenue, Cambridge, MA 02139 USA (e-mail: uferner,medard@mit.edu). E. Soljanin is at Bell Labs, Alcatel-Lucent, 600 Mountain Av., Murray Hill, NJ 07974 (e-mail: emina@research.bell-labs.com).

- Using this model, we derive blocking probabilities of SANs with uncoded storage;
- We analyze an NCS scheme using a similar framework to random linear network coding (RLNC) and show that NCS performs better than uncoded storage; and
- We present a simple energy consumption model for SANs to illustrate how reductions in blocking probability translate into energy savings.

To increase the speed of data distribution and download speeds, the use of network coding has garnered significant attention, primarily in peer-to-peer networks [9]–[11]. In systems closer to SANs such as distributed storage, network coding has been proposed for data repair [12]–[14]. However to the authors' knowledge, little work has considered the application of network coding in SANs for blocking probability improvements. Network designs for the explicit management of network energy have focused on full DCs or content distribution networks [15], [16], as opposed to individual SANs.

The remainder of this paper is organized as follows. Section II provides preliminary material, including the SAN energy model and the SAN service model. Section III develops a theoretical analysis of NCS and its effects on SAN energy. Section IV discusses new directions for research and concludes the paper.

## II. Preliminaries

Firstly, this section describes the typical energy breakdown for a single SAN. Secondly, the SAN video-streaming based service model used later in the paper is described.

### A. Data Center Energy Model

The energy use of a SAN is decomposed into the storage of data and the energy used in the transmission of that stored data. Storage energy in the SAN is consumed by (i) servers;[1] (ii) storage units; and (iii) auxiliary units such as cooling, office lighting, and load balancers. Transmission energy is defined as the energy consumed by all active routers and switches between SAN drives and users' terminals. A typical energy breakdown is depicted in Fig. 1.

We compute the average energy $E(x)$ of a SAN, composed of servers and routers [15] using

$$E(x) = xp\left(n\,\gamma_s + h\gamma_r\right), \qquad (1)$$

[1] Modern DCs can have anywhere between 1 and 300 000 servers. [2]

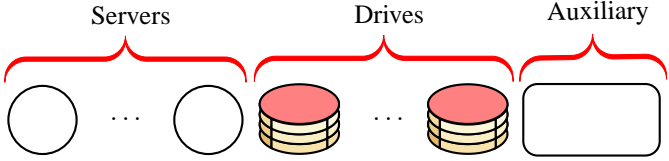

| Servers | | Drives | | Auxiliary |
|---|---|---|---|---|
| CPU, RAM | 87W | Per drive | 15W | |
| Other | 211W | | | |
| Total/Server | 325W | Total/Server | ≈90W | Total/Server 340W |

Fig. 1: A typical SAN is composed of multiple servers, drives, and auxiliary components such as cooling and lighting [15], [17], [18]. A typical breakdown of the energy consumption of these units is depicted. The iconography used throughout this paper is also shown.

TABLE I: Typical network parameters in the SAN energy consumption model for (1).

| Parameter name | Notation | Typical value |
|---|---|---|
| Number of hops from a user to a DC | $h$ | 14 |
| Router energy/bit | $\gamma_r$ | 150 J/Gb |
| Server energy/bit | $\gamma_s$ | 572.3 J/Gb |
| PUE | $p$ | 2 |

where $x$ is the communication load in Gb/s and $p$ is a constant power usage effectiveness (PUE).[2] Parameters $\gamma_s$ and $\gamma_r$ are the server and router energy per bit, respectively. Note that $\gamma_s$ and $\gamma_r$ are both measured in W / Gb/s = J/Gb. Finally, $h$ is the average number of hops from a user to the DC and $n$ is the number of servers in the SAN.

As per Fig. 1, the direct power requirements of storage units are small in comparison to servers and auxiliary units and (1) ignores direct storage unit power consumption. Critically however, although the power requirements of storage units are small, reducing the number of storage units correspondingly reduces the number of servers required to manage those storage units. It is through this coupling of server and storage requirements that reducing storage requirements can reduce SAN energy requirements.

It is instructive to illustrate typical parameters values in (1). Industry standards for PUE $p$ range from 1.09 to 3, depending on the size and sophistication of the systems of interest [1], [19]. On the server energy per bit $\gamma_s$, numerous studies have sought to measure this experimentally by estimating the total power consumed by a single server and applying affine load-to-power models. We use experimentally derived estimates of $\gamma_s = 572.3$ and $\gamma_r = 150$ [15]. See Table I for a summary; these will be used in Section III to calculate total SAN energy consumption.

### B. Storage Area Network Service Model

The analysis of SAN energy usage requires an understanding of the availability of requested content over time, where the available content is located, as well as the service time for that content. The modeled hardware components that service user read requests are load balancers, servers, I/O buses and storage drives. We describe the connectivity of these components as well as the service models for each.

Upon arrival of a user's read request, that request traverses a path through the following hardware components: The request arrives at a load balancer and is then forwarded onto a subset of servers. Those servers attempt to access connected drives to read out and transfer the requested content back to the user. Let $\{S_i\}_{i=1}^n$ be the set of $n$ SAN servers, and $\{D_{u,j}\}_{j=1}^{m_u}$ be the $m_u$ drives connected to $S_u$. Define $v = \max\{m_1, \ldots, m_n\}$ as the maximum number of drives to which a single server can be connected.

Any individual drive can only concurrently access a limited number of read requests—this restriction is particularly pronounced in the case of high-definition (HD) video—so each drive has an I/O bus with access bandwidth $B$ bits/second [20], and a download request requires a streaming and fixed bandwidth of $b$ bits/second. Component connectivity is shown in Fig. 2.

We use the following notation for files and chunks therein. Let drives in the SAN collectively store a file library $\mathcal{F} = \{f_1, \ldots, f_F\}$, where $f_i$ is the $i$th file, and there are $F$ files stored in total. Each file $f_i$ is decomposed into equal-sized chunks, and all files are of the same size. Note that, in the NCS scheme, chunks are the units across which coding is performed. Typical chunk sizes for video files in protocols such as HTTP Live Streaming (HLS) are on the order of a few seconds of playback [21], although this is dependent on various codec parameters. The SAN stores $W$ copies of each file. We order the $T$ chunks of file $f_i$ in time, by $f_i^{(k)} = \{f_{i,1}^{(k)}, \ldots, f_{i,T}^{(k)}\}$, where $f_{i,j}^{(k)}$ is the $k$th copy of the $j$th ordered chunk of file $i$.

We allow striping of file chunks across multiple drives. Striping is a technique in which chunks from the same file are systematically distributed across multiple disks [20] to speed up read times. An example of a common striping standard is the RAID0 standard. In particular, a server striping file $f_i^{(k)}$ may read sequential chunks from the same file in a round-robin fashion among multiple drives. We make the following assumptions about file layout throughout the SAN, and the striping of content:

- File are not striped among servers, i.e., each file copy $f_i^{(k)}$ is managed by only a single server and chunks of $f_i^{(k)}$ are not split across drives connected to different servers;
- Define $s$ as the number of drives across which each file is striped; if a file is striped across $s$ drives, we refer to it as an *s-striped file*; and
- The contents of each drive that stores a portion of an $s$-striped file is in the form $f_{i,j}^{(k)}, f_{i,j+s}^{(k)}, f_{i,j+2s}^{(k)}, \ldots$, and define these chunks as the *jth stripe-set for file $f_i$*.

As an example, consider a SAN in which $f_i^{(1)}$ is striped across three drives. A server connected to all drives would read chunks in the following order: (1) $f_{i,1}^{(1)}$ from $D_{1,1}$; (2) $f_{i,2}^{(1)}$ from $D_{1,2}$; (3) $f_{i,3}^{(1)}$ from $D_{1,3}$; (4) $f_{i,4}^{(1)}$ from $D_{1,1}$, and so on.

We model user read requests as a set of independent Poisson processes. In particular, we invoke the Kleinrock independence assumption [22] and model arriving read requests for each chunk $f_{i,j}$ as a Poisson process with arrival rate $n\lambda$, which is independent of other chunk read request arrivals. Upon a request arrival, the load balancer randomly assigns the request to some server $S_u$ with uniform distribution. (This splits the incoming Poisson process and each server sees requests at rate $\lambda$.) Server $S_u$ then requests the relevant chunks from its connected drives $\{D_{u,j}\}_{j=1}^{m_u}$. Since a download request requires a streaming and fixed bandwidth of $b$ bits/second then if the requested file is $s$-striped, each drive I/O bus will require access bandwidth of size $b/s$. See Fig. 2 for an illustration. In addition, the ratio $sB/b$ is the *number of access bandwidth slots* that each active drive has available to service read arrivals. Once a particular drive's I/O access bandwidth is allocated, that drive has an average service time of $1/\mu$ seconds to service that chunk read request.

---
[2]The auxiliary services of a SAN are captured by the industry standard "power usage effectiveness" multiplier, defined as the ratio of the total energy consumed over the energy consumed by the IT equipment such as servers, external storage, and internal routers and switches [15].

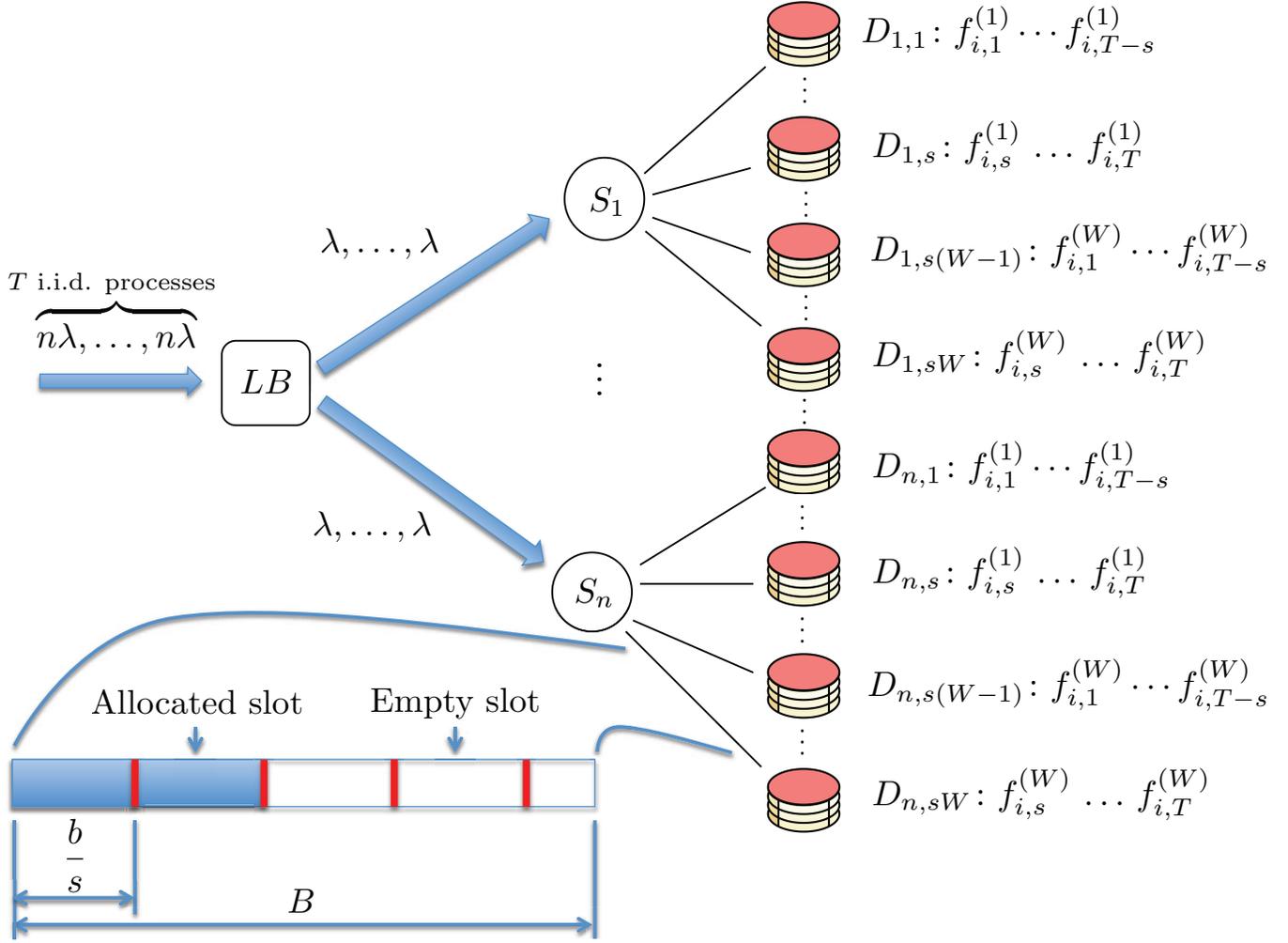

Fig. 2: Hardware components in the SAN service model. The load balancer is denoted by the auxiliary node $LB$, server $m$ by $S_m$, and drive $(m, j)$ by $D_{m,j}$. Each drive is connected to a single server through an I/O bus with access bandwidth $B$ b/s. Each chunk read request arrives at and is processed by $LB$ prior to being forwarded to some server $S_i$, and then a set of drives. File $f_i$ is $s$-striped and each server has access to $W$ copies of each file chunk, each on different drives. In this connectivity layout, drive contents are shown to the right of each colon.

A drive can only accept a new request from a server if it has sufficient I/O access bandwidth $b/s$ available at the instant the read request arrives. If instead all access bandwidth slots are currently allocated, then that request is rejected or *blocked* by that drive. If a request is accepted by a drive then that drive's controller has determined it can meet the various read timing guarantees for that request and a bandwidth slot is allocated. Internally, each drive has a disk controller queue for requests and some service distribution governing read request times [23]–[27]. However, thanks to the internal disk controller's management of request timing guarantees, all accepted request reads begin service immediately from the perspective of the server.[3] If a drive currently cannot accept new read requests from servers we say that drive is in a *blocked state*. If no useful information relevant to a download request at server $S_u$ can be serviced from any connected drives upon that requests arrival,

---
[3]The full service distribution for modern drives such as SATA drives is dependent on numerous drive model specific parameters including proprietary queue scheduling algorithms, disk mechanism seek times, cylinder switching time, and block segment sizes [25].

then that server blocks or rejects that read request.

## III. Energy consumption of NCS and UCS

This section begins by determining the SAN blocking probability in a UCS scheme as a function of the number of drives and the striping number. The NCS scheme is then described, after which the corresponding blocking probabilities are determined. The energy consumption functions of UCS and NCS schemes are then contrasted and compared.

In the UCS scheme, without loss of generality, set the library $\mathcal{F} = \{f_i\}$ to be a single $s$-striped file with $W$ copies of each chunk in the SAN. If no drive contains more than one copy of a single chunk, then $sW \leq m_u$, $\forall u \in \{1, \ldots, n\}$. Assume that all chunks have uniform read arrival rates so $\lambda = \lambda_j \, \forall j \in \{1, \ldots, T\}$. As discussed, the path traversed by each Poisson process arrival is shown in Fig. 2, and once a chunk read is accepted by a drive, that drive takes an average time of $1/\mu$ to read the request.

We model the blocking probability of this system as follows. File $f_i$ is *blocked* if there exists at least one chunk in $f_i$ that is in a blocked state. Chunk $f_{i,j}$ is available if there exists a drive that contains it and has an available access bandwidth slot. An $s$-striped drive that holds a single stripe set may service requests from either $\lceil T/s \rceil$ or $\lfloor T/s \rfloor$ different chunks, depending on the length of the stripe set. We merge all read requests for chunks from $j$th stripe set of a file copy into a single Poisson process with arrival rate

$$\lambda \lfloor T/s \rfloor + \mathbb{I}(j \leq T \bmod s), \quad (2)$$

where $j \in \{1, \ldots, s\}$ is a drive index containing the $j$th stripe-set, and $\mathbb{I}$ is the indicator function. For each file copy, there will be $T \bmod s$ drives with rate $\lceil T/s \rceil$ and $s - T \bmod s$ with rate $\lfloor T/s \rfloor$. We map each access bandwidth slot onto a single independent service unit from an $M/G/K^U/K^U$ queue, see for instance [28], in which each queue has $K^U$ service units. There are $W$ copies of any stripe set on different drives, so our $M/G/K^U/K^U$ queue has

$$K^U = \lfloor sBW/b \rfloor \quad (3)$$

independent service units for the $j$th stripe set. This mapping is depicted in Fig. 3. The $M$ denotes that the arrival process is Poisson; $G$ denotes a general service distribution with average rate $\mu$; and $K^U$ denotes the total number of service units in the system, as well as the maximum number of active service requests after which incoming requests are discarded [28].

The blocking probability $P_b^{(j)}$ of the $j$th stripe set queue is given by the well-studied *Erlang B* blocking formula,

$$P_b^{(j)} = \begin{cases} \frac{(\rho \lceil T/s \rceil)^{K^U}}{e^{(\rho \lceil T/s \rceil)} \Gamma(1+K^U, \rho \lceil T/s \rceil)}, & j \leq T \bmod s \\ \frac{(\rho \lfloor T/s \rfloor)^{K^U}}{e^{(\rho \lfloor T/s \rfloor)} \Gamma(1+K^U, \rho \lfloor T/s \rfloor)} & \text{else} \end{cases}$$

where $\rho = \lambda/\mu$ and $\Gamma$ is the upper incomplete Gamma function. The probability that a chunk is available is equal to $1 - P_b^{(j)}$. The probability that $f_i$ is blocked $P_b^U$, i.e., the probability that not all chunks are available, is then given by

$$P_b^U = 1 - \left(1 - \frac{(\rho \lceil T/s \rceil)^{K^U}}{e^{\rho \lceil T/s \rceil} \Gamma(1+K^U, \rho \lceil T/s \rceil)}\right)^{T \bmod s}$$
$$\times \left(1 - \frac{(\rho \lfloor T/s \rfloor)^{K^U}}{e^{\rho \lfloor T/s \rfloor} \Gamma(1+K^U, \rho \lfloor T/s \rfloor)}\right)^{s - T \bmod s}. \quad (4)$$

### A. NCS Design

We now describe our NCS scheme and compute the corresponding blocking probability. NCS is equivalent to UCS except that we replace each chunk $f_{i,j}^{(k)}$ from the SAN with a coded chunk $c_{i,j}^{(k)}$. To allow for video streaming applications,

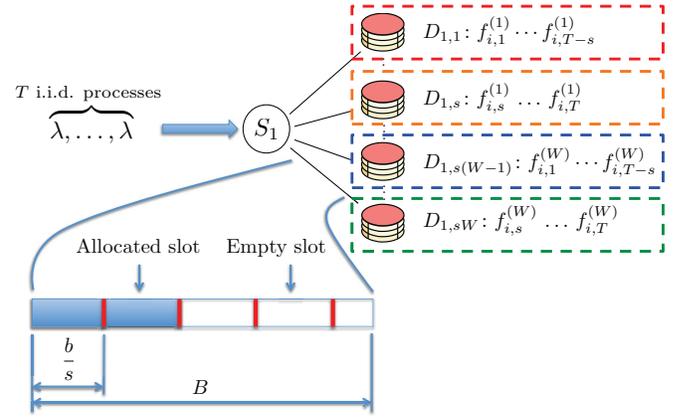

(a) An example SAN in which a single server has access to $W$ copies of file $f_i$. Each connected drive has total access bandwidth $B$, and each slot takes bandwidth $b/s$. Hence, each connected drive has $\lfloor sB/b \rfloor$ available access bandwidth slots and employs the UCS scheme. Using UCS, the queue mapping for this architecture is shown below in Fig. 3(b). For simplicity of illustration, we assume integrality of $T/s$.

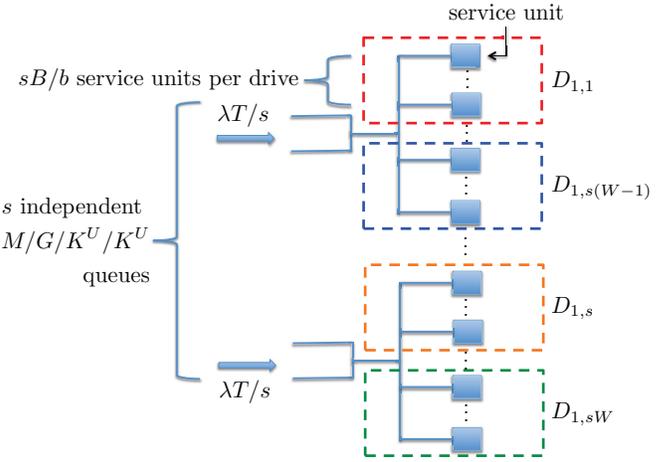

(b) The equivalent $M/G/K^U/K^U$ mapping from Fig. 3(a). For simplicity of illustration, we assume integrality of $T/s$.

Fig. 3: An example mapping from a network architecture with a single server to an $M/G/K^U/K^U$ queue in the UCS scheme. Note that $s$ independent queues exist in this mapping. For simplicity of illustration, we assume $T \bmod s = 0$.

we wish to allow a user to receive, decode, and begin playing chunks at the beginning of a file prior to having received the entire file. Coded chunks are constructed as follows. We divide each file into $L$ equal-sized block windows or generations, each containing $r$ chunks. Owing to striping, we constrain $r \leq s$ and $s/r \in \mathbb{N}^+$. (There will be no performance gain from network coding if there is coding across chunks on the same drive, and coding across chunks on the same drive must exist if $r > s$.) Let $\mathcal{B}_{i,l}$ be the $l$th block window/generation, where $\mathcal{B}_{i,l}$ is a subset of file $f_i$'s chunk indices and $\mathcal{B}_{i,l}$ is disjoint from all other block windows. See Fig. 4(a) for an illustration.

Coded chunk $c_{i,j}^{(k)}$, $j \in \mathcal{B}_{i,l}$, is a linear combination of all uncoded chunks in the same block window that contains $f_{i,j}$,

$$c_{i,j}^{(k)} = \sum_{p \in B_{i,l}} \alpha_{p,j}^{(k)} f_{i,p}^{(k)} \qquad (5)$$

where $\alpha_{p,j}^{(k)}$ is a column vector of coding coefficients drawn from a finite field $\mathbb{F}_q$ of size $q$ [29], and where we treat $f_{i,p}^{(k)}$ as a row vector of elements from $\mathbb{F}_q$. We assign coding coefficients that compose each $\alpha_{p,j}^{(k)}$ with uniform distribution from $\mathbb{F}_q$, mirroring RLNC [30], in which the random coefficients are continuously cycled. In this scheme, coded chunk $c_{i,j}$ now provides the user with partial information on all chunks in its block window. Note that coefficients are randomly chosen across both the chunk index $j$ as well as the copy $k$. Similarly to [12], when a read request arrives for a coded chunk, the relevant drive transmits both $c_{i,j}^{(k)}$ as well as the corresponding coefficients $\{\alpha_{p,j}^{(k)}\}$.

In the NCS scheme, the blocking probability is determined as follows. Similar to UCS, we merge the independent Poisson arrival processes for uncoded chunks $\{f_{i,j}: j \in B_{i,l}\}$ into a Poisson process with arrival rate either $r\lambda \lceil T/s \rceil$ or $r\lambda \lfloor T/s \rfloor$, depending on the stripe-set length. This process can be interpreted as requests for any coded chunk that has an innovative degree of freedom in the $l$th block window. See Fig. 4 for an example mapping from a hardware architecture to a queue in which $W = 1$. Generalizing such an architecture, the request rates for an innovative chunk in the $l$th block window are again mapped to an equivalent $M/G/K^C/K^C$ queue with parameter

$$K^C = rK^U, \qquad (6)$$

and so the blocking probability $P_b^{(jC)}$ for each coded stripe set $M/G/K^C/K^C$ queue is given by

$$P_b^{(jC)} = \begin{cases} \frac{(r\rho\lceil T/s \rceil)^{K^C}}{e^{r\rho\lceil T/s \rceil}\Gamma(1+K^C, r\rho\lceil T/s \rceil)}, & j \leq (T \bmod s)/r \\ \frac{(r\rho\lfloor T/s \rfloor)^{K^C}}{e^{r\rho\lfloor T/s \rfloor}\Gamma(1+K^C, r\rho\lfloor T/s \rfloor)}, & \text{else}. \end{cases}$$

The constraint $s/r \in \mathbb{N}^+$ ensures that the queuing model has $s/r$ independent queuing systems and that no intra-drive coding exists, in a similar fashion to the UCS scheme. This

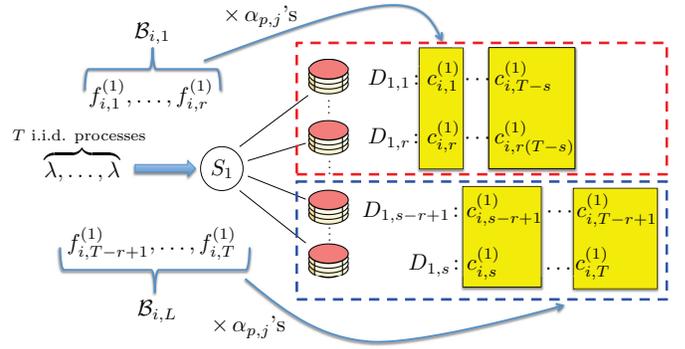

(a) An example of a single server system that only has access to a single copy of file $f_i$. This depiction with $W = 1$ is in contrast to Fig. 3 and is only for visual simplicity. Chunks are coded using NCS, and those in the same highlighted block are composed of coded chunks from the same block window.

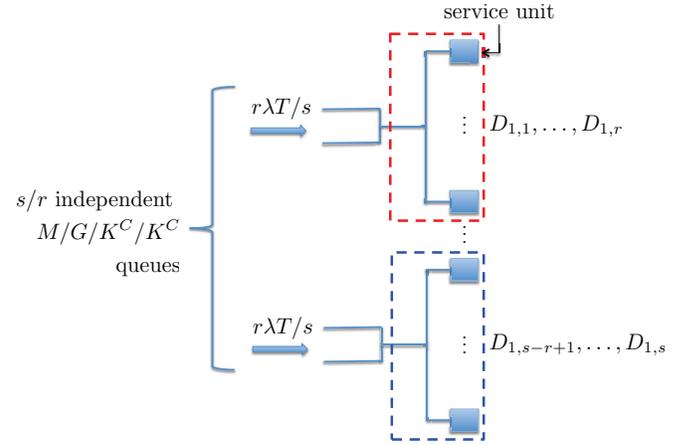

(b) A queue mapping from Fig. 4(a).

Fig. 4: An example mapping from a single server hardware architecture with a single copy $W = 1$ of file $f_i$ to an $M/G/K^C/K^C$ queue in an NCS in blocks of $r$ chunks. In Fig. 4(a), file chunks have been coded using NCS, and those in the same highlighted block are composed of coded chunks from the same block window as per (5). For simplicity of illustration, we assume integrality of $T/s$.

implies the NCS blocking probability $P_b^C$ is given by

$$P_b^C = 1 - \left(1 - \frac{(r\rho\lceil T/s \rceil)^{K^C}}{e^{r\rho\lceil T/s \rceil}\Gamma(1+K^C, r\rho\lceil T/s \rceil)}\right)^{s/r} \\ \times \left(1 - \frac{(r\rho\lfloor T/s \rfloor)^{K^C}}{e^{r\rho\lfloor T/s \rfloor}\Gamma(1+K^C, r\rho\lfloor T/s \rfloor)}\right)^{\frac{s}{r} - \lfloor \frac{T \bmod s}{r} \rfloor}.$$

(7)

### B. UCS and NCS Comparison

We now compare the blocking probabilities and energy efficiencies of NCS and UCS. Fig. 5 plots (4) and (7) as a function of $W$ for three different stripe-rates $s = 2, 4, 8$, which we refer to as low, medium, and high stripe-rates, respectively. We set the number of chunks to $T = 150$ to approximate a short movie trailer if chunks are divided up using a protocol such as HLS. Finally, the number of videos that each drive can concurrently service $B/b$ is set to 2.

As the stripe-rate increases, the benefit of the NCS scheme over UCS become more apparent. In particular, assuming a target quality of service (QOS) of $P_b = 10^{-8}$, in the low stripe-rate scenario the number of file copies required has approximately a 20% savings. In contrast, in the high stripe-rate scenario the number of file copies required has approximately a 50% savings.

### C. Super-chunks

We now sharpen the distinction between NCS and existing chunk-layout schemes. It is reasonable to ask, if coding chunks across drives provides blocking probability gains, then does amalgamating chunks together in other formats provide similar gains? We consider an example of amalgamating sequential chunks $(f_{i,j}, \ldots, f_{i,j+l})$ into *super-chunks* and compare the blocking probability performance of that scheme to the aforementioned UCS scheme. We adapt the UCS scheme to use super-chunks as follows. Consider the super chunk file layout depicted in Fig. 6. If the chunk size is increased $l$ times in comparison to normal chunks, then each super-chunk will take $l$ times longer to be read. Keeping the number of chunks or superchunks on each drive constant for comparison purposes, then in the super-chunk scheme we have an $s/l$-striped file as opposed to an $s$-striped file. Consider independent arrival processes for each super-chunk each with rate $l\lambda$ and that the entire super-chunk is returned upon request. The traffic intensity for the super-chunk model is given by

$$l\lambda T/s/\mu/l = l^2 \rho T/s. \tag{8}$$

In addition, owing to striping over a smaller number of drives we have less bandwidth available to service incoming requests. This drops the number of service units in the queuing system to

$$K^{SC} = \lceil K^U/l \rceil. \tag{9}$$

Assuming integrality of $s/l$, the final blocking probability for the super chunk scheme is then given by

$$P_b^{SC} = 1 - \left(1 - \frac{(l^2\rho T/s)^{\lceil K^U/l \rceil}}{e^{l^2\rho T/s}\Gamma(1 + \lceil K^U/l \rceil, l^2\rho T/s)}\right)^{s/l}. \tag{10}$$

Fig. 7 provides an example plot comparing the blocking probabilities of the normal chunk layout to the super-chunk file layout, with $l = 2$. The performance of the super-chunk file layout is significantly degraded compared to the normal chunk layout since the rate of the system has increased and the number of service units available to service those requests has decreased.

Alternative super-chunk layouts exist where, for example, the memory per drive is kept constant as opposed to the number of chunks per drive. In such a layout the load of the system increases by a factor of $l$ and the number of service units remains the same; again the blocking probability is higher than the normal file layout. The performance gain of NCS is tightly coupled to the increase in the number of available service units to service chunk read requests. Different file layout schemes such as super-chunks can certainly allow system designers to rearrange queue statistics by merging queues, but it is *the inter-drive information coupling and the increase in available service units that follows from it which is a key differentiator between NCS and such existing placement strategies.*

### D. Energy and Financial Implications

We now estimate the potential energy savings for a SAN that uses NCS using the model outlined in Section II-A. The two primary terms in the energy coefficient in (1) are the $n\gamma_s$ and $h\gamma_r$ terms. Although the storage of data with RLNC would likely reduce communication requirements and hence reduce $h$, we assume there are no differences in $h$ in NCS and UCS. An estimate of the order of magnitude of energy savings follows. As an illustration, assume that each server manages up to twelve drives, $v = 12$ and that each drive stores 2 terabytes (TB) of data: Hence every 24 TB of data saved allows one server to be switched off. Referring to Fig. 1 and Table I, a server directly consumes 325 W and with a typical PUE of 2, each switched off server yields a 650 W/24 TB, i.e., 27 W/TB, saving.

A percentage energy saving illustration is estimated using (1). In an enterprise SAN server energy consumption dominates router consumption, so

$$\begin{aligned} E(x) &= xp \left( n\,\gamma_s + h\gamma_r \right) \\ &\approx xp\,n\gamma_s \,, \end{aligned} \tag{11}$$

and with a target $P_b = 10^{-8}$, we estimate that the *energy consumption is reduced by 20–50%* as one moves from a low to high striping-rate regime.

We now convert these energy savings into SAN operating expenditure financial savings. A typical annual energy cost for an enterprise-size SAN is USD$3M [4] and over a typical ten year lifetime, savings can be estimated using a discounted cashflow model to compute the net present value (NPV) [31]:

$$NPV = \sum_{i=1}^{10} \frac{CF_i}{(1+WACC)^i} \,, \tag{12}$$

where $NPV$ is the net present value, $CF_i$ is the cash flow in year $i$, and $WACC$ is the weighted cost of capital for the industry of interest given by [32]

$$WACC = \frac{D}{D+E}r_D + \frac{E}{D+E}r_E \,. \tag{13}$$

In this case, $D$, $E$, $r_D$, and $r_E$ are the average debt, equity, cost of debt, and cost of equity in the DC industry, respectively. We use the average ratio of values as per Table II to compute an *NPV savings of USD$4.04–10.1M*. This significant financial saving stems from the fact that small duplication savings are significantly magnified in enterprise level SANs owing to large server and drive numbers.

## IV. Discussion & Conclusions

This paper has characterized the energy savings of NCS in SANs. We introduced a mapping technique to analyze SANs as independent $M/G/K/K$ queues and derived the blocking probability for UCS schemes. These UCS schemes were then contrasted with NCS, in which potential energy savings of 20–50% were illustrated. It was demonstrated that blocking probability gains scale well with striping number and that NCS operates differently from chunk amalgamation.

TABLE II: Estimated finance performance parameters for the DC industry, taken from publicly available financial market data for the U.S. communication equipment industry [33].

| Parameter | Estimated value |
|---|---|
| $D/(D+E)$ | 0.55 |
| $r_D$ | 0.07 |
| $r_E$ | 0.09 |

The presented blocking probability gains of NCS over UCS are dependent on the stripe rate and the Kleinrock independence assumption of arrivals. The Kleinrock assumption for the independence of incoming chunk read requests requires significant traffic mixing and moderate-to-heavy traffic loads [22]. This models reality most closely in large SANs with sufficient traffic-mixing from different users and with traffic loads such as those found in enterprise-level DCs. In contrast, in smaller SANs, such as those found in closet DCs, the number of users and the traffic load are smaller and the correlated effects of arrivals between chunks in the same file will become more important.

Given the Kleinrock assumption, results show that blocking probability improvements scale well with striping rates, i.e., as striping rates increase with NCS, the number of required file copies decreases. This may motivate the exploration of very high stripe rates in certain systems for which blocking probability metrics are of particular concern. Although we do not specifically advocate for very high stripe-rates, the overall benefits of striping will remain in use in some form as long as there is a demand and supply mismatch between I/O. Over the last two decades, application bitrate demand growth has continued to outpace increases in drives' I/O growth; as long as this trend continues the striping paradigm is likely to remain and the benefits of very high stripe-rates may require further exploration.

Future work includes extending analysis of NCS to correlated arrivals between chunks, to inter-SAN architectures and to full CDN architectures. In addition, the implementation of NCS in a testbed could provide additional compelling insights.

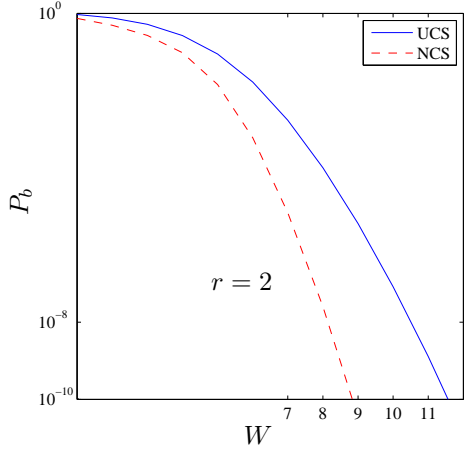

(a) The effect of NCS on duplication requirements in an example single SAN on the blocking probability $P_b$ with a low stripe-rate in a system with low load. In this setup the stripe-rate is set to $s = 2$, the number of chunks is $T = 150$, $B/b = 2$ and $\rho = 0.2$.

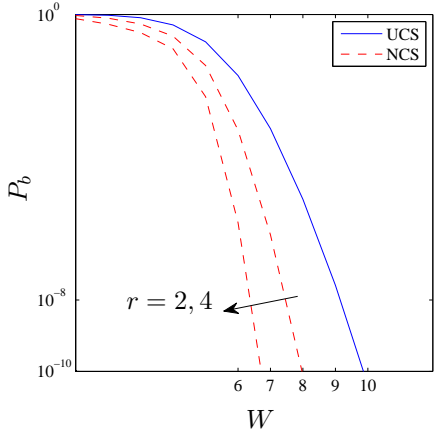

(b) The effect of NCS on duplication requirements in a single SAN on the blocking probability $P_b$ with a medium stripe-rate in a system with heavy load. In this setup the stripe-rate is set to $s = 4$, the number of chunks is $T = 150$, $B/b = 2$ and $\rho = 0.9$.

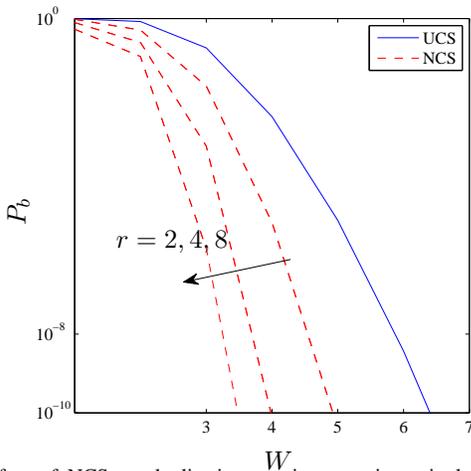

(c) The effect of NCS on duplication requirements in a single SAN on the blocking probability $P_b$ with a high stripe-rate in a system with heavy load. In this setup the stripe-rate is set to $s = 8$, the number of chunks is $T = 150$, $B/b = 1$ and $\rho = 0.9$.

Fig. 5: The effect of NCS on duplication requirements as a function of blocking probability under various stripe-rates and system loads.

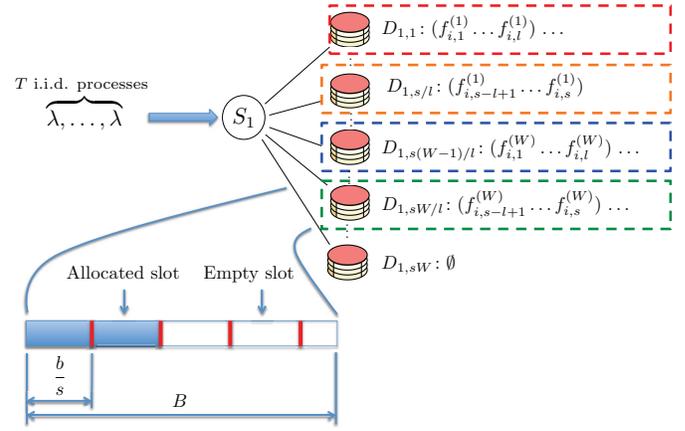

Fig. 6: An example of a super chunk file layout. Each connected drive has $sB/b$ available access bandwidth slots and employs the UCS scheme with super chunks. Each super chunk is an amalgamated set of $l$ chunks.

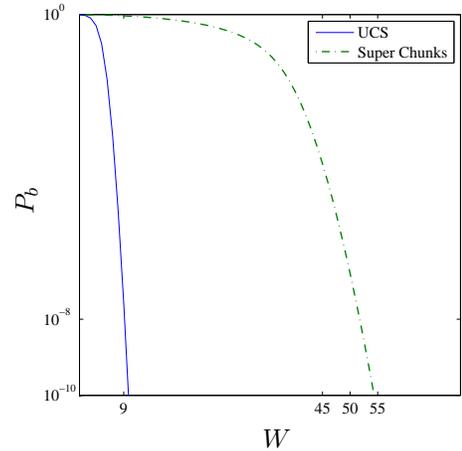

Fig. 7: A comparison of blocking probability for the UCS and the super-chunk scheme. The number of chunks is set to $T = 150$, $s = 4$, $B/b = 2$ and $\rho = 0.9$. The size of each super-chunk is two amalgamated chunks so $l = 2$.